\documentclass[preprint,12pt]{elsarticle}

\usepackage{amssymb}
\usepackage{amsthm}
\usepackage{amsmath}
\usepackage{enumitem}
\usepackage{ctable}
\usepackage{algorithm}
\usepackage{regexpatch}
\usepackage{algpseudocode}
\usepackage{multirow}
\usepackage{mathtools}
\usepackage{mathrsfs}
\usepackage{amsfonts}

\usepackage[top=1.25in, bottom=1.2in, left=1.25in, right=1.25in]{geometry}

\usepackage{listings}
\usepackage{xcolor}   

\definecolor{darkgreen}{rgb}{0,0.6,0}
\definecolor{mauve}{rgb}{0.88, 0.69, 1.0}
\definecolor{operamauve}{rgb}{0.72, 0.52, 0.65}
\definecolor{mauvetaupe}{rgb}{0.57, 0.37, 0.43}
\definecolor{darkgray}{rgb}{0.66, 0.66, 0.66}

\lstdefinestyle{my_Common_list}{
  basicstyle=\small\color{blue}\ttfamily,
  numbers=left,
  numberstyle=\tiny\color{gray},  
  stepnumber=1,                   
  numbersep=5pt,                  
  backgroundcolor=\color{white}, 
  showspaces=false,               
  showstringspaces=false,         
  showtabs=false,                 
  rulecolor=\color{black},        
  tabsize=1,                 
  captionpos=b,              
  breaklines=true,                    
  breakatwhitespace=false,     
  title=\lstname,                       
  keywordstyle=\color{blue},          
  commentstyle=\color{darkgreen},       
  stringstyle=\color{mauvetaupe},         
  escapeinside={\%*}{*)},            
  morekeywords={*,...},               
  belowskip=-1.5em,
  escapeinside={(*}{*)}  
}

\lstdefinestyle{my_Common_small}{
  style=my_Common_list,
}

\lstdefinestyle{XML} { %
  language=XML,                
  style=my_Common_list,
}

\lstdefinestyle{CPP} { %
  language=c++,                
  style=my_Common_list,
}

\lstdefinestyle{bash} { %
  language=bash,               
  style=my_Common_list,
}

\lstdefinestyle{Fortran}
{
    style=my_Common_list,
    language={Fortran},
    alsolanguage={[LaTeX]TeX},
}

\makeatletter
\renewcommand{\ALG@beginalgorithmic}{\footnotesize}
\xpatchcmd{\algorithmic}{\itemsep\z@}{\itemsep=0.15ex}{}{}
\makeatother
\algrenewcommand\algorithmicindent{1.0em}%
\algrenewcommand\alglinenumber[1]{\scriptsize #1:}
\algrenewcommand{\algorithmiccomment}[1]{\hfill$!$ #1}

\usepackage{listings}
\lstdefinestyle{Common}
{
    basicstyle=\scriptsize\ttfamily\null,
    numbersep=1em,
    frame=single,
    framesep=\fboxsep,
    framerule=\fboxrule,
    xleftmargin=\dimexpr\fboxsep+\fboxrule,
    xrightmargin=\dimexpr\fboxsep+\fboxrule,
    breaklines=true,
    breakindent=0pt,
    tabsize=5,
    columns=flexible,
    showstringspaces=false,
    captionpos=b,
    abovecaptionskip=0.5\smallskipamount,   
}

\lstdefinestyle{Fortran}
{
    style=Common,
    language={Fortran},
    alsolanguage={[LaTeX]TeX},
    morekeywords=
    {
    },
}

\newcounter{bla}

\begin{document}

\begin{frontmatter}



\title{\textsc{ElVibRot-MPI}: parallel quantum dynamics with Smolyak algorithm for general molecular simulation}


\author[a]{Ahai Chen}
\ead{achenphysics@gmail.com}
\address[a]{Universit{\'e} Paris-Saclay, UVSQ, CNRS, CEA, Maison de la Simulation, 91191, Gif-sur-Yvette, France}

\author[b,c]{Andr\'e Nauts}
\ead{andre.nauts@uclouvain.be}
\address[b]{Institute of Condensed Matter and Nanosciences (NAPS), Universit\'e Catholique de Louvain, Louvain-la-Neuve, Belgium}

\author[c]{David Lauvergnat}
\ead{david.lauvergnat@universite-paris-saclay.fr}
\address[c]{Universit{\'e} Paris-Saclay, CNRS, Institut de Chimie Physique, UMR-CNRS 8000, 91405 Orsay, France}

%

\begin{abstract}
A parallelized quantum dynamics package using the Smolyak algorithm for general molecular simulation is introduced in this work. The program has no limitation of the Hamiltonian form and provides high flexibility on the simulation setup to adapt to different problems. Taking advantage of the Smolyak sparse grids formula, the simulation could be performed with high accuracy, and in the meantime, impressive parallel efficiency.  The capability of the simulation could be up to tens of degrees of freedom. The implementation of the algorithm and the package usage are introduced, followed by typical examples and code test results. 
\end{abstract}

\begin{keyword}
quantum dynamics; Smolyak algorithm; sparse grids; parallel computation
\end{keyword}

\end{frontmatter}



{\bf PROGRAM SUMMARY}

\begin{small}
\noindent
{\em Program Title: \textsc{ElVibRot-MPI}}                                          \\
{\em Licensing provisions: LGPL}                                   \\
{\em Programming language: Fortran 90 \& some Fortran 2003, Fortran 77}                                   \\
{\em Nature of problem: Solving the Schr{\"o}dinger equation for general quantum dynamics simulation.}\\
{\em Solution method: Smolyak sparse-grids algorithm, curvilinear coordinate}\\
{\em Additional comments including Restrictions and Unusual features: Applicable for general quantum simulation up to tens of degrees of freedom. No limitation of Hamiltonian form. No built-in limitation of degrees of freedom.}\\
   \\

%
%

\end{small}

\section{Introduction}
\label{}
The rapid growth of effects on quantum dynamics simulation in the past decades has got insights into the fundamental properties of molecular dynamics, triggered a new level of the understanding of chemical reaction, laser-matter interaction, etc.~\cite{McCullough1969_QD_initial,review_Tully2012_QMD,review_Curchod2018_QMD}. The rigorous quantum simulation, in accord with the increasing complexity of considered systems, is demanding more computation resources and, in another perspective, more efficient algorithms. In particular, various methods have been proposed to overcome the difficulties lies in the exponential growth of computational demand with the increasing of the system dimension, the ``curse of dimension", in the standard direct-product basis scheme. 

Some well-known methods includes the quantum diffusion Monte Carlo (DMC)~\cite{Reynolds1990_DMC,Martin1991_DMC,Kosztin1996_DMC,Clary2001_DMC,Philip2016_DMC}, the Feynman path integral molecular dynamics approaches~\cite{Feynman1948_path_int,Beutier2014_FPI}, the vibrational self-consistent field~\cite{Benoit2006_VSCF,Christiansen2007_VSCF}, the multi-configuration time-dependent Hartree (MCTDH) ~\cite{review_Beck2000_MCTDH,book_Meyer2009_MCTDH} method and its extended versions~\cite{book_gonzalez2020_QD}, to mention but a few. The former two methods avoid expending the wave function on a basis set, being applicable for getting properties of large systems. The MCTDH method expands the wave function as a summary of Hartree products with single-particle functions. It has established a high efficiency in wave-packet propagation, but normally requests a transformation of the potential for specific problems. Besides, another choice for dealing with the curse of dimensionality is the Smolyak algorithm, proposed by Smolyak in 1963~\cite{Smolyak1963_initial,Gradinaru2008_smolyak_math,Lasser2020_smolyak_math}. The method proposes a sparse grid to efficiently represent high-dimensional grids. It has been very successfully applied for the optimal algorithms for high-dimensional integration~\cite{Novak1996_smolyak,Petras2003_smolyak}, the solution of differential equations~\cite{Bungartz2004_smolyak_PDE}, etc. The application of Smolyak method in quantum simulation has been proposed recently~\cite{Avila2009_smolyak_initial,Avila2011_smolyak}. It has been applied for getting exchange-correlation energies for density functional theory~\cite{Rodriguez2008_smolyak_DFT,Rodriguez2008p_smolyak_DFT} and computing the vibrational spectra of molecules up to 12 degrees of freedom~\cite{Avila2009_smolyak_initial,Avila2011_smolyak,Avila2015_smolyak,David2014_smolyak,David2019_smolyak}. Taking advantage of the Smolyak formula (see next section for details), the parallelization of Smolyak algorithm could further improve this application.

In this work, we introduce a package ``\textsc{ElVibRot-MPI}" for general quantum simulations using a parallelized Smolyak algorithm. The program is designed to be highly flexible in the simulation setup and to fit different types of machines. It is available for the simulation of general molecules up to tens of degrees of freedom on moderate-scale computational nodes. The paper is organized as follows. In the second section, we introduce the method employed in the program; Then, the parallel implementation of the Smolyak algorithm is described in the third section; In the fourth section, the code usage is introduced, including the parameters, the installation and the running of the program. Finally, in section 5, the benchmark check and efficiency tests are presented with discussions, followed by a conclusion in section 6. 

\section{Method}
The core of a quantum dynamics simulation lies in solving the Schr{\"o}dinger equation 
\begin{eqnarray} 
i\hbar\frac{\partial}{\partial t}|\Psi\rangle=\hat{H}|\Psi\rangle,
\end{eqnarray}
where $\hat{H}$ is the Hamiltonian of the system. Generally, in the direct-product scheme, the 
wave function can be expanded on the direct-product  representation 
\begin{eqnarray}  \label{eq_dir_rep}
S^{dp}=B^1 \otimes B^2\otimes \cdots\otimes B^n
\end{eqnarray}
as
\begin{eqnarray}  \label{eq_dir_expension}
|\Psi\rangle=\sum_{k_1=1}^{N^{b^1}}...\sum_{k_n=1}^{N^{b^n}}\psi^{B^1,...,B^n}_{k_1,k_2,...,k_n}|b^{1}_{k_1}\rangle|b^{2}_{k_2}\rangle\cdots|b^{n}_{k_n}\rangle,
\end{eqnarray}
where $B^i\equiv\{b^{i}_{1},...,b^{i}_{k_i},...,b^{i}_{N^{b^i}}\}$ $(i\in [1,n])$ is the primitive basis sets. $\psi^{B^1,...,B^n}_{k_1,k_2,...,k_n}$ is the coefficient of wave function expansion. 
Express the basis  $|b^{i}_{k_i}\rangle$ as $b^{i}_{k_i}({\bf Q}^i)$ in coordinates representation, the wave function reads
 \begin{eqnarray} 
 \psi({\bf Q}^1,...,{\bf Q}^n)=\sum_{k_1=1}^{N^{b^1}}...\sum_{k_n=1}^{N^{b^n}}\psi^{B^1,...,B^n}_{k_1,k_2,...,k_n}b^{1}_{k_1}({\bf Q}^1)b^{2}_{k_2}({\bf Q}^2)\cdots b^{n}_{k_n}({\bf Q}^n).
 \end{eqnarray}
Reversely, the coefficient $\psi^{B^1,...,B^n}_{k_1,k_2,...,k_n}$ could be obtained from the wave function as 
 \begin{eqnarray}
 \psi^{B^1,...,B^n}_{k_1,k_2,...,k_n}
 &=&\langle b^{1}_{k_1}b^{2}_{k_2}\cdots b^{n}_{k_n}|\psi\rangle\nonumber\\
 &=&\int d\tau b^{1}_{k_1}({\bf Q}^1)b^{2}_{k_2}({\bf Q}^2)\cdots b^{n}_{k_n}({\bf Q}^n) \psi({\bf Q}^1,{\bf Q}^2,...,{\bf Q}^n)
  \end{eqnarray}
 Numerically, the integral could be performed on certain grid. The basis set $B^i$ is associated to a grid $G^i\equiv\{Q^i_1,...,Q^i_{u_i},...,Q^i_{N^{q^i}}\}$ with relevant weights $\{w_1,...,w_i,...,w_{N^{q^i}}\}$. Then we obtain
 \begin{eqnarray}\label{eq_g_to_b}
 \psi^{B^1,...,B^n}_{k_1,k_2,...,k_n}&=&\sum_{u_1=1}^{N^{q^1}}\cdots\sum_{u_n=1}^{N^{q^n}}\left[w_1b^{1}_{k_1}({\bf Q}^1_{u_1})w_2b^{2}_{k_2}({\bf Q}^2_{u_2})\cdots w_nb^{n}_{k_n}({\bf Q}^n_{u_n})\right.\nonumber\\
 &&\left.\cdot\psi({\bf Q}^1_{u_1},{\bf Q}^2_{u_2},...,{\bf Q}^n_{u_n})\right].
 \end{eqnarray}
 where 
 \begin{eqnarray}\label{eq_b_to_g}
\psi^{G^1,...,G^n}_{u_1,u_2,...,u_n}
&\equiv&\psi({\bf Q}^1_{u_1},{\bf Q}^2_{u_2},...,{\bf Q}^n_{u_n})\nonumber\\
&=&\sum_{k_1=1}^{N^{b^1}}\cdots\sum_{k_n=1}^{N^{b^n}}\psi^{B^1,...,B^n}_{k_1,k_2,...,k_n}b^{1}_{k_1}({\bf Q}^1_{u_1})b^{2}_{k_2}({\bf Q}^2_{u_2})\cdots b^{n}_{k_n}({\bf Q}^n_{u_n}).
 \end{eqnarray}
 is the wave function on grid.
 
Following this procedure we could obtain a numerical accurate solution of the Schr{\"o}dinger equation. However, it would be extremely expensive to compute when increasing the system dimension $n$. In the Smolyak method~\cite{Smolyak1963_initial,Gradinaru2008_smolyak_math,Avila2009_smolyak_initial,David2019_smolyak,Lasser2020_smolyak_math}, instead of the direct-product form Eq.\ref{eq_dir_rep}, we describes the system with a Smolyak representation 
\begin{eqnarray}
S_{L}=\sum_{\mathscr{R}(l)}D_{l_1l_2...l_n}S^1_{l_1}\otimes S^2_{l_2}\otimes \cdots\otimes S^n_{l_n},
\end{eqnarray}
with the restriction $\mathscr{R}(l)$: $L-n+1\leq\sum_{i=1}^{n}l_i\leq L$.
$D_{l_1l_2...l_n}=(-1)^{L-|l|}C_{n-1}^{L-|l|}$. $C_{n-1}^{L-|l|}$ is the binomial coefficient. $L$ is a constant that controls the approximation level, which is defined both for basis ($L^B$) and grid ($L^G$).  $S^i_{l_i}$ ($i\in[1,n]$) is the restricted basis set $\{b^{i}_{1},b^{i}_{2},...,b^{i}_{N^{b^i}_{l_i}}\}$ or grid set $\{Q^i_1,Q^i_2,...,Q^i_{N^{q^i}_{l_i}}\} $.
The relation between $N_{l_i}$ (for both $N^{b_i}_{l_i}$ and $N^{q_i}_{l_i}$) and $l_i$ could be in principle an arbitrarily increasing integer sequence function, e.g. $N_{l_i}=A_i+B_il_i$ ($A_i\geq0$, $B_i\geq1$, $A_i,B_i\in\mathbb{Z}$). However, a well-chosen function could be essential for specific basis sets~\cite{Avila2011_smolyak,Andre2018_smolyak,David2018_smolyak_clathrate}. 

The wave function in $S_L$ could be expressed as 
\begin{eqnarray} \label{eq_wf_smolyak}
|\Psi\rangle&=&\sum_{\mathscr{R}(l)}D_{l_1...l_n}|\psi^{l_1...l_n}\rangle.
\end{eqnarray}
where the Smolyak term $|\psi^{l_1l_2...l_n}\rangle$ is a smaller direct-products. On basis, it reads
\begin{eqnarray}  \label{eq_dir_expension_st}
|\psi^{B^1_{l_1}B^2_{l_2}...B^n_{l_n}}\rangle=\sum_{k_1=1}^{N_{l_1}^{b^1}}...\sum_{k_n=1}^{N_{l_n}^{b^n}}\psi^{B_{l_1}^1,...,B_{l_n}^n}_{k_1,k_2,...,k_n}|b^{1}_{k_1}\rangle|b^{2}_{k_2}\rangle\cdots|b^{n}_{k_n}\rangle,
\end{eqnarray}
Namely, the wave function is reduced to the summary of $N^{st}=\sum_{\mathscr{R}(l)}$ weighted smaller ``Smolyak wave functions". The total number of possible basis functions could be obtained with 
\begin{eqnarray}
N_s=\sum_{\mathscr{R}(l)}\left(\prod_{i=1}^{n}N_{l_i}\right).
\end{eqnarray} 
Furthermore, the wave function could be presented as a mixture of  Smolyak term on basis and grid presentation as either 
\begin{eqnarray}
|\Psi\rangle=~\sum_{\mathclap{\resizebox{0.130\textwidth}{!}{$L-n+1\leq|l|\leq L$}}}~~D_{l_1l_2...l_n}|\psi^{B^1_{l_1}B^2_{l_2}...B^n_{l_n}}\rangle+~\sum_{\mathclap{\resizebox{0.130\textwidth}{!}{$\substack{L-n+1\leq|l'|\leq L\\|l|'\neq|l|}$}}}~~D_{l_1l_2...l_n}|\psi^{G^1_{l_1}G^2_{l_2}...G^n_{l_n}}\rangle
\end{eqnarray}
or
\begin{eqnarray}
|\Psi\rangle=~\sum_{\mathclap{\resizebox{0.130\textwidth}{!}{$L-n+1\leq|l|\leq L$}}}~~D_{l_1l_2...l_n}|\psi^{B^1_{l_1}B^2_{l_2}...B^i_{l_i}G^{i+1}_{l_{i+1}}...G^n_{l_n}}\rangle,
\end{eqnarray}
or the mixture of both. For the transformation between the basis and the grid respresenation for a certain Smolyak term, following Eq.\ref{eq_g_to_b}, \ref{eq_b_to_g}, we get
\begin{eqnarray}\label{eq_transfer_g_to_b}
\psi^{B^1_{l_1}B^2_{l_2}...B^i_{l_i}G^{i+1}_{l_{i+1}}...G^n_{l_n}}=\sum_{u_i=1}^{N^{q^i}}w_ib^{i}_{k_i}({\bf Q}^i_{u_i})\psi^{B^1_{l_1}B^2_{l_2}...G^i_{l_i}G^{i+1}_{l_{i+1}}...G^n_{l_n}}
\end{eqnarray}
for a transfer from $G^i_{l_i}$ to $B^i_{l_i}$, and reversely
\begin{eqnarray}\label{eq_transfer_b_to_g}
\psi^{B^1_{l_1}B^2_{l_2}...G^i_{l_i}G^{i+1}_{l_{i+1}}...G^n_{l_n}}=\sum_{k_i=1}^{N^{b^i}}b^{i}_{k_i}({\bf Q}^i_{u_i})\psi^{B^1_{l_1}B^2_{l_2}...B^i_{l_i}G^{i+1}_{l_{i+1}}...G^n_{l_n}}.
\end{eqnarray}
It is  an important feature for an efficient numerical operator action.

As a result, for a general operator $\hat{O}$, the operator action in Smolyak algorithm takes the form
\begin{eqnarray}\label{eq_action_smolyak}
\hat{O}|\Psi\rangle&=&\sum_{\mathscr{R}(l)}D_{l_1l_2...l_n}\hat{O}|\psi^{l_1l_2...l_n}\rangle.
\end{eqnarray}
Therefore, the Schr{\"o}dinger equation reads
\begin{eqnarray}
i\hbar\frac{\partial}{\partial t}\sum_{\mathscr{R}(l)}D_{l_1l_2...l_n}|\psi^{l_1l_2...l_n}\rangle
=\sum_{\mathscr{R}(l)}D_{l_1l_2...l_n}\hat{H}|\psi^{l_1l_2...l_n}\rangle.
\end{eqnarray}
This formula has a good structure for parallel computation. The accuracy of the method depends on the $L$ and $N_{l_i}$, namely, the relation between $N_{l_i}$ and $l_i$.  
Higher $L$ increases the coupling of different dimensions.  

\begin{figure}   [ht!]
\begin{center}
\centering
\includegraphics[width=0.470\textwidth]{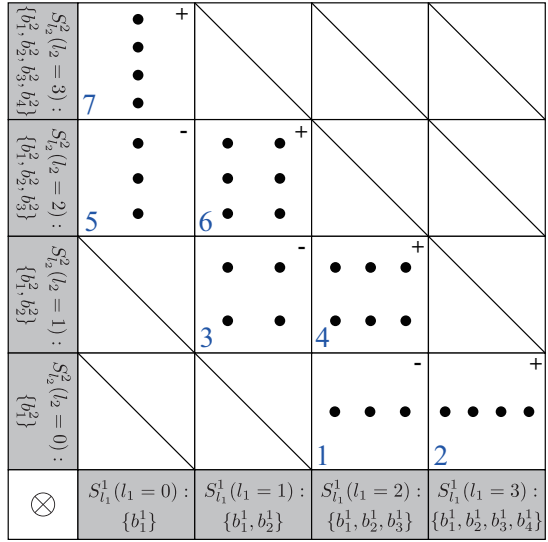}
\caption[]{An example of Smolyak terms for the basis sets in a 2D system with $L^B=3$, $N^{b^i}_{l_i}=1+l_i$. The Smolyak terms are indicated with blue numbers. The basis functions in each Smolyak term are shown as black dots. The value of $D_{l_1l_2...l_n}$ is shown on the top right corner of each Smolyak term.} \label{fig_example_smolyak}
\end{center}
\end{figure}

As a typical example, we show in Fig.\ref{fig_example_smolyak} the Smoyak terms for the basis sets in a 2-dimensional system with $L=3$ and $N^{b^i}_{l_i}=1+l_i$. The primary basis sets for different $l_i$ are shown on the labels. The possible Smoyak terms are indicated as blue numbers with the contained basis functions shown as black dots. The value of $D_{l_1l_2...l_n}$ is shown in the top right corner. The selected Smolyak terms gathers in the lower triangle region, requesting much fewer basis functions. Moreover, some basis functions could present in several different Smolyak terms. Thus, in the 7 Smolyak terms, the total number of selected basis functions is 30, or 10 without duplication.  

Furthermore, in Table \ref{table_num_terms}, we present the number of the basis functions required in Smolyak algorithm when $N^{i}_{l_i}=1+l_i$ at different $L$ for systems with 12 or 120 degrees of freedom. The case with ($N_s$) or without duplication ($N_{cs}$) of basis functions, as well as their ratio, are provided.
The result of the direct-product scheme $N_{dp}=(L+1)^n$ is shown for comparison. The ratio between $N_s$ and $N_{cs}$ is increasing with the degrees of freedom and $L$, indicating more duplicated basis functions in larger $L$ or larger systems. The number of the basis functions required is still affordable at 120 degrees of freedom in Smolyak method.

\begin{table} [ht!]
\begin{center}
\begin{tabular}[t]{cc|cccccl}
\specialrule{0.10em}{0em}{0em}
        &    & L=2  &  L=3     & L=4       & L=5        & L=6 \\
\hline
\multicolumn{1}{c|}{\multirow{4}{*}{12D}}  & $N_{dp}$ & 4096 &  531441   & 16777216   & 244140625 & 2176782336  \\
\multicolumn{1}{c|}{}                                  & $N_s$     & 325 &  2925   & 20475   &   118755  & 593775  \\
\multicolumn{1}{c|}{}                                  & $N_{cs}$ & 91   &  455    & 1820    &     6188     & 18564     \\
\cline{2-7}
\multicolumn{1}{c|}{}                                  & $N_s/N_{cs}$         & 3.57 & 6.43 & 11.25 & 19.19 & 31.99     \\            
\hline
\multicolumn{1}{c|}{\multirow{4}{*}{120D}} & $N_{dp}$& $10^{36}$ & $10^{57}$ & $10^{72}$ & $10^{83}$ & $10^{93}$ \\
\multicolumn{1}{c|}{}                                  & $N_s$    &  29161 & 2362041 & 144084501 & 7060140549 & 289465762509  \\
\multicolumn{1}{c|}{}                                  & $N_{cs}$ & 7381    & 302621   &  9381251    & 234531275   & 4925156775      \\
\cline{2-7}
\multicolumn{1}{c|}{}                                  & $N_s/N_{cs}$          & 3.95 & 7.81   & 15.36   & 30.10  & 58.77   \\            
\specialrule{0.10em}{0em}{0em}
\end{tabular}
\caption{
The number of basis functions required in a 12- or 120-degrees of freedom system in the direct-product scheme $N_{dp}$ and the Smolyak method with $N^{i}_{l_i}=1+l_i$ at different $L$, where $N_s$ and $N_{cs}$ present the results with or without the duplication of the basis functions in Smolayk terms. The ratio between $N_s$ and $N_{cs}$ is also shown.
}\label{table_num_terms}
\end{center}
\end{table}

\section{Implementation}
To implement the Smolyak algorithm, we first introduce three objects:  
\begin{enumerate}[nolistsep]
\item Smolyak representation on basis. It consists of a large number of Smolyak terms $|\psi^{B^1_{l_1}B^2_{l_2}...B^n_{l_n}}\rangle$. These terms do not need to present in the memory at the same time, thus is not actually memory consuming. 

\item compact Smolyak representation on basis $|\psi^{B^1B^2...B^n}\rangle$. The compact basis is obtained by removing the duplicated basis functions in all the Smolyak terms to save memory in the simulation.

 \item Smolyak representation on grid. It contains the Smolyak terms on the grid $|\psi^{G^1_{l_1}G^2_{l_2}...G^n_{l_n}}\rangle$. It is the representation where we perform operator action.
\end{enumerate}
The transform between $|\psi^{B^1_{l_1}B^2_{l_2}...B^n_{l_n}}\rangle$ and $|\psi^{G^1_{l_1}G^2_{l_2}...G^n_{l_n}}\rangle$ depends on Eq.\ref{eq_transfer_g_to_b} and \ref{eq_transfer_b_to_g}. However, to bridge $|\psi^{B^1B^2...B^n}\rangle$ and all $|\psi^{B^1_{l_1}B^2_{l_2}...B^n_{l_n}}\rangle$, an extra mapping table $M^B$ is required. For example, the $M^B$ for the case in Fig.\ref{fig_example_smolyak} is shown in Table.\ref{table_mapping}. The Smolyak terms, the  basis functions, and the compact basis functions are indexed as $i^{st}\in[1,N_{st}]$, $i^s\in[1,N_s]$, and $i^{cs}\in[1,N_{cs}]$, respectively. Therefore, the key of the parallelization lies on the balance of the memory of $|\psi^{B^1B^2...B^n}\rangle$, $M^{B}$, $|\psi^{B^1_{l_1}B^2_{l_2}...B^n_{l_n}}\rangle$ and the MPI communication time.

\begin{table} [ht!]
\begin{center}
\resizebox{0.990\textwidth}{!}{
{\setlength{\tabcolsep}{1.7pt}
\begin{tabular}[t]{c|*{32}{c}}
\specialrule{0.10em}{0em}{0em}
$i^{st}$ &  \multicolumn{3}{c|}{\multirow{1}{*}{1}}  &  \multicolumn{4}{c|}{\multirow{1}{*}{2}} &  \multicolumn{4}{c|}{\multirow{1}{*}{3}} & \multicolumn{6}{c|}{\multirow{1}{*}{4}} &  \multicolumn{3}{c|}{\multirow{1}{*}{5}} & \multicolumn{6}{c|}{\multirow{1}{*}{6}} &   \multicolumn{4}{c}{\multirow{1}{*}{7}} \\
\hline
 $\{l_1$,$l_2\}$ & \multicolumn{3}{c|}{\multirow{1}{*}{\{2,0\}}} &   \multicolumn{4}{c|}{\multirow{1}{*}{\{3,0\}}} & \multicolumn{4}{c|}{\multirow{1}{*}{\{1,1\}}} & \multicolumn{6}{c|}{\multirow{1}{*}{\{2,1\}}} & \multicolumn{3}{c|}{\multirow{1}{*}{\{0,2\}}} & \multicolumn{6}{c|}{\multirow{1}{*}{\{1,2\}}} & \multicolumn{4}{c}{\multirow{1}{*}{\{0,3\}}} \\
 \hline
$i^s$ & 1 & 2 & 3 & 4 & 5 & 6 & 7 & 8 & 9 &10 &11 &12 &13 &14 &15 &16 &17 &18 &19 &20 &21 &22 &23 &24 &25 &26 &27 &28 &29 &30 \\
$i^{cs}$ & 1 & 5 & 8 & 1 & 5 & 8 &10 & 1 & 5 & 2 & 6 & 1 & 5 & 8 & 2 & 6 & 9 & 1 & 2 & 3 & 1 & 5 & 2 & 6 & 3 & 7 & 1 & 2 & 3 & 4 \\
\specialrule{0.10em}{0em}{0.1em}
\end{tabular}
}}
\begin{tabular}[t]{c|*{10}c}
\specialrule{0.10em}{0.1em}{0em}
compact basis & $b^1_1b^2_1$ & $b^1_1b^2_2$  & $b^1_1b^2_3$  & $b^1_1b^2_4$  & $b^1_2b^2_1$  & $b^1_2b^2_2$  & $b^1_2b^2_3$  & $b^1_3b^2_1$  & $b^1_3b^2_2$  & $b^1_4b^2_1$\\
\hline
$i^{cs}$ & 1 & 2 & 3 & 4 & 5 & 6 & 7 & 8 & 9 & 10 \\
\specialrule{0.10em}{0em}{0em}
\end{tabular}
\caption{Top table shows the mapping table $M^B$ for the case in Fig.\ref{fig_example_smolyak}. The Smolyak terms, the basis functions, and the compact basis are indexed as $i^{st}$, $i^s$ and $i^{sc}$, respectively. The indexing of basis function in compact representation is shown in the bottom table. }\label{table_mapping}
\end{center}
\end{table}

With the three objects, the steps for an operator action in Smolyak method (see Eq.\ref{eq_action_smolyak}) could be described as follows:
\begin{enumerate}[nolistsep]
\item extract one Smolyak term from the compact basis with $M^B$.
\item transfer from the basis to the grid representation according to Eq.\ref{eq_transfer_b_to_g}.
\item perform operator action to get new Smolyak term on grid.
\item transfer back to basis representation with Eq.\ref{eq_transfer_g_to_b}.
\item compress and add the new Smolyak term to new compact basis $|\widetilde{\psi}^{B^1B^2...B^n}\rangle$.
\item repeat the above steps for all $N_{st}$ Smolyak terms to get $|\widetilde{\psi}^{B^1B^2...B^n}\rangle$.
\end{enumerate}
The algorithm is shown in Alg.\ref{algorithm_one_smolyak}. The operator action is performed for one Smolyak term each time as the standard direct-product scheme, thus not memory consuming. 

\begin{algorithm}[ht!]
\caption{Operator action of one Smolyak term}\label{algorithm_one_smolyak}
\begin{algorithmic}[1]
\Procedure{Action\_Smolyak\_term}{$\hat{O}$, $|\psi^{B^1B^2...B^n}\rangle, |\widetilde{\psi}^{B^1B^2...B^n}\rangle$,{$i^{st}$}} \Comment{$i^{st}\in[1,N_{st}]$}
    \State $\{l_1l_2...l_n\}\gets i^{st}$
    \State $|\psi^{B^1_{l_1}B^2_{l_2}...B^n_{l_n}}\rangle \xleftarrow{M^{B_{l_1l_2...l_n}}} |\psi^{B^1B^2...B^n}\rangle$ 
    \State $|\psi^{G^1_{l_1}G^2_{l_2}...G^n_{l_n}}\rangle  \gets  |\psi^{B^1_{l_1}B^2_{l_2}...B^n_{l_n}}\rangle$ \Comment{Eq.\ref{eq_transfer_b_to_g}}
    \State $|\widetilde{\psi}^{G^1_{l_1}G^2_{l_2}...G^n_{l_n}}\rangle=\hat{O}|\psi^{G^1_{l_1}G^2_{l_2}...G^n_{l_n}}\rangle$  \Comment{conventional operator action}
    \State $|\widetilde{\psi}^{B^1_{l_1}B^2_{l_2}...B^n_{l_n}}\rangle\gets |\widetilde{\psi}^{G^1_{l_1}G^2_{l_2}...G^n_{l_n}}\rangle$ \Comment{Eq.\ref{eq_transfer_g_to_b}}
    \State $|\widetilde{\psi}^{B^1B^2...B^n}\rangle \gets \sum D_{l_1l_2...l_n}|\widetilde{\psi}^{B^1_{l_1}B^2_{l_2}...B^n_{l_n}}\rangle$ \Comment{the contribution from one Smolyak term}
    \State \textbf{return} $|\widetilde{\psi}^{B^1B^2...B^n}\rangle$
\EndProcedure
\end{algorithmic}
\end{algorithm}

\begin{figure}[ht!]
\begin{center}
\centering
\includegraphics[width=0.80\textwidth]{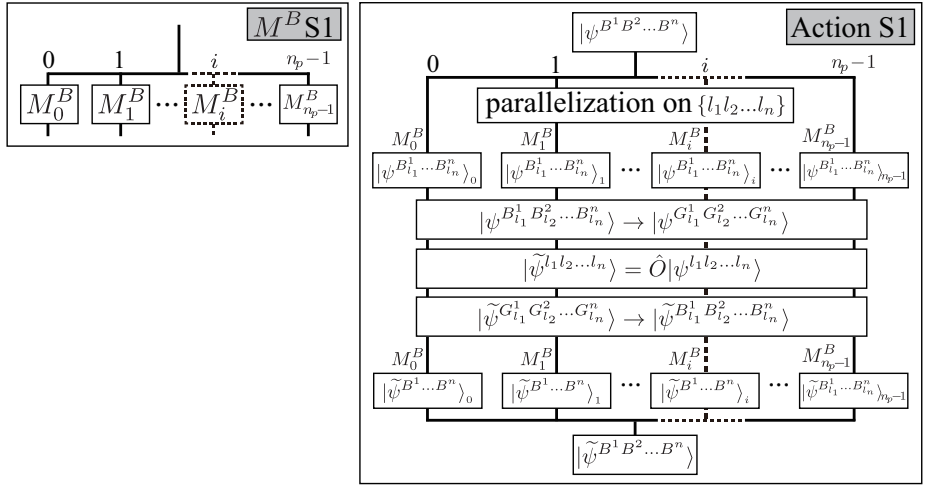}
\caption[]{The panels illustrate the flowchart of  mapping table ($M^B$ S1) and the operator action (Action S1) in  MPI schemes 1. $n_{p}$ indicates the number of available processors. See the main text for more details of the notations. }\label{fig_action_S1}
\end{center}
\end{figure}

\begin{figure}[ht!]
\begin{center}
\centering
\includegraphics[width=0.80\textwidth]{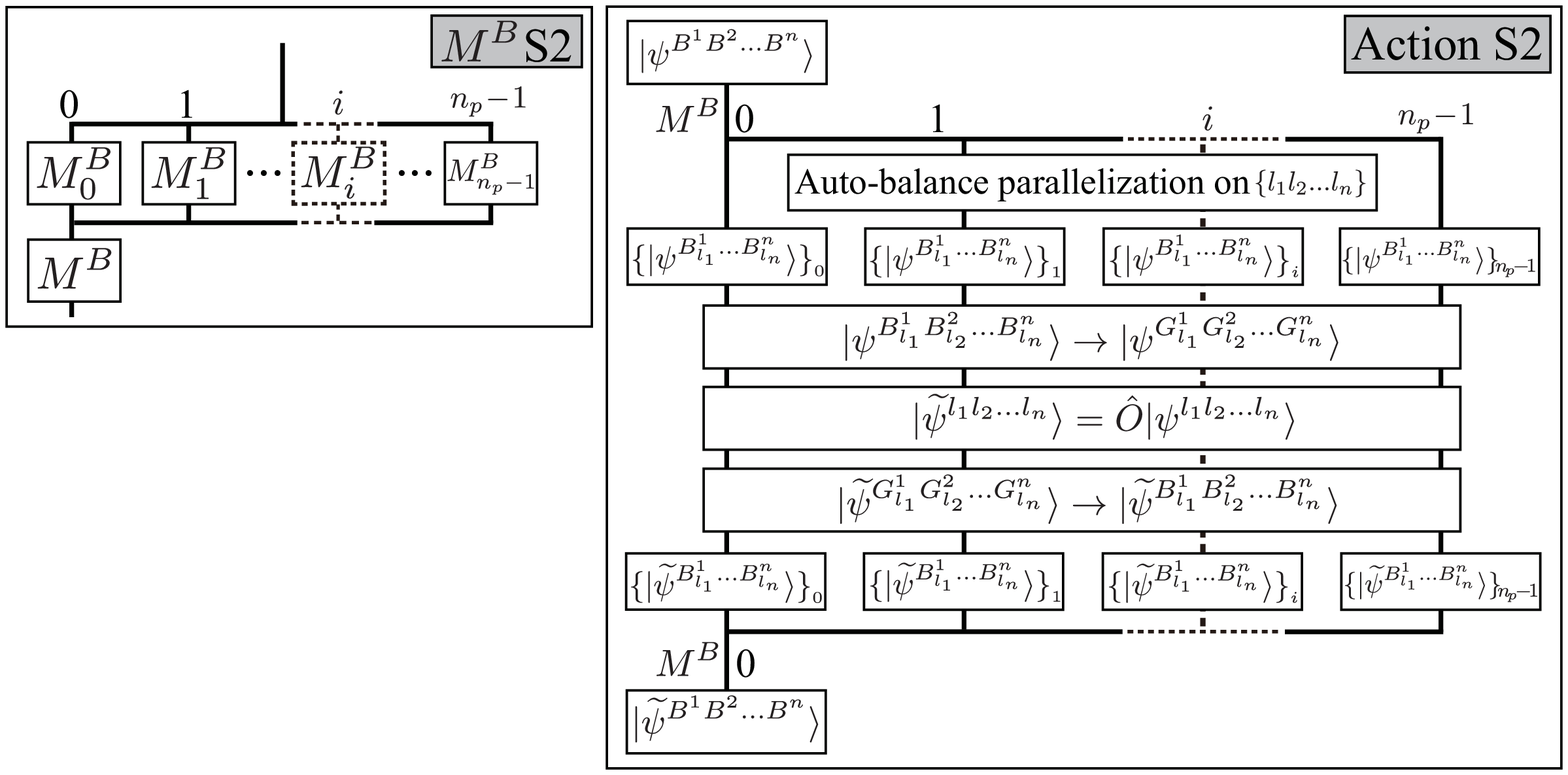}
\caption[]{Same as Fig.\ref{fig_action_S1} but for scheme 2.}\label{fig_action_S2}
\end{center}
\end{figure}

The parallelization of operator action depends on Eq.\ref{eq_action_smolyak}. For different molecules and parameters $L$ and $N^{b^i}_{l_i}$, the memory consumption of  $|\psi^{B^1_{l_1}B^2_{l_2}...B^n_{l_n}}\rangle$, $|\psi^{B^1B^2...B^n}\rangle$ and  $M^{B}$ diverse.  Therefore, different strategies could be embedded. In principle, there are two basic MPI schemes according to the memory consumption of $|\psi^{B^1B^2...B^n}\rangle$ and $M^{B}$: 

In the first scheme,  each processor performs part of the operator action with identical compact basis $|\psi^{B^1B^2...B^n}\rangle$ and the required portion of the mapping table $M_p^{B}$ ($p=0...n_p-1$), where $p$ indexes the processors, $n_p$ is the number of available processors.
The algorithm is shown in Alg.\ref{algorithm_smolyak_MPI1}. As an illustration, the basic flowcharts of the scheme for the operator action and the related mapping table are presented in Fig.\ref{fig_action_S1}, indicated as Action S1 and $M^B$ S1, respectively.
This scheme is of the best performance when there is large enough memory assigned to each processor, i.e. $\{|\psi^{B^1B^2...B^n}\rangle\}^{\mathcal{M}}+\{M_p^{B}\}^{\mathcal{M}}<\{p\}^{\mathcal{M}}$, where $\{\cdots\}^\mathcal{M}$ denotes the memory reqiured by $|\psi^{B^1B^2...B^n}\rangle$, $M_p^{B}$ and the available memory of the processor $p$, respectively. 
In the second scheme, $|\psi^{B^1B^2...B^n}\rangle$  and $M^{B}$ are kept only on a master processor, the Smolyak terms are extracted and distributed to the other processors for the calculation.  It increases the MPI communication time but reduces the overall memory consumption. The relevant algorithm and flowchart are shown in Alg.\ref{algorithm_smolyak_MPI2} and Fig\ref{fig_action_S2}, respectively.

\begin{algorithm}[ht!]
\caption{MPI Operator action with Smolyak method - basic scheme 1}\label{algorithm_smolyak_MPI1}
\begin{algorithmic}[1]
\Procedure{Action\_Smolyak\_MPI1}{$\hat{O}$, $|\psi^{B^1B^2...B^n}\rangle,|\widetilde{\psi}^{B^1B^2...B^n}\rangle$}
  \State [$i^{st}_{1_p}$, $i^{st}_{2_p})~\xleftarrow{p} N_{st}$   \Comment{assign Smolyak terms to processor $p$} 
  \For {$i^{st}$=$i_{1_p}^{st}$, $i_{2_p}^{st}$} 
    \State \Call{Action\_Smolyak\_term}{$\hat{O}$, $|\psi^{B^1B^2...B^n}\rangle_p, |\widetilde{\psi}^{B^1B^2...B^n}\rangle_p$, $i^{st}$}
  \EndFor
  \State  $|\widetilde{\psi}^{B^1B^2...B^n}\rangle\gets$  {\bf MPI sum} $|\widetilde{\psi}^{B^1B^2...B^n}\rangle_{p}$ 
  \State {\bf MPI boardcast} $|\widetilde{\psi}^{B^1B^2...B^n}\rangle$
  \State \textbf{return} $|\widetilde{\psi}^{B^1B^2...B^n}\rangle$
\EndProcedure
\end{algorithmic}
\end{algorithm}

\begin{algorithm}[ht!]
\caption{MPI Operator action with Smolyak method - basic scheme 2}\label{algorithm_smolyak_MPI2}
\begin{algorithmic}[1]
\Procedure{Action\_Smolyak\_MPI2}{$\hat{O}$, $|\psi^{B^1B^2...B^n}\rangle,|\widetilde{\psi}^{B^1B^2...B^n}\rangle$}
  \If{on master processor}  \Comment{$p$=0}
   \State [$i^{st}_{1_p}$, $i^{st}_{2_p}$)$~\xleftarrow{p} N_{st}$ 
    \State $\{|\psi^{B^1_{l_1}B^2_{l_2}...B^n_{l_n}}\rangle\}_{p} \xleftarrow{M^B} |\psi^{B^1B^2...B^n}\rangle$  \Comment{$\{...\}_{p'}$ contains all $\{l_1l_2...l_n\}\rightarrow i^{st}\in[i^{st}_{1_{p'}}, i^{st}_{2_{p'}})$}
    \State {\bf MPI send}  $\{|\psi^{B^1_{l_1}B^2_{l_2}...B^n_{l_n}}\rangle\}_{p'}$ to processor $p'$  \Comment{$p'\in[1,n_p-1]$}
    \State \Call{$\{|\widetilde{\psi}^{B^1_{l_1}B^2_{l_2}...B^n_{l_n}}\rangle\}_0 \gets$ Action\_Smolyak\_basis}{$\hat{O}$,$\{|\psi^{B^1_{l_1}B^2_{l_2}...B^n_{l_n}}\rangle\}_0$, $i^{st}_{1_0}$, $i^{st}_{2_0}$}
    \State   {\bf MPI receive} $|\widetilde{\psi}^{B^1_{l_1}B^2_{l_2}...B^n_{l_n}}\rangle_{p'}$ from processor $p'$ \Comment{$p'\in[1,n_p-1]$}
    \State $|\widetilde{\psi}^{B^1B^2...B^n}\rangle \gets \sum D_{l_1l_2...l_n}|\widetilde{\psi}^{B^1_{l_1}B^2_{l_2}...B^n_{l_n}}\rangle$ 
 \Else
    \State {\bf MPI receive} $\{|\psi^{B^1_{l_1}B^2_{l_2}...B^n_{l_n}}\rangle\}_{p'}$    
    \State \Call{$\{|\widetilde{\psi}^{B^1B^2...B^n}\rangle\}_{p'} \gets$ Action\_Smolyak\_basis}{$\hat{O}$, $\{|\psi^{B^1_{l_1}B^2_{l_2}...B^n_{l_n}}\rangle\}_{p'}$, $i^{st}_{1_{p'}}$, $i^{st}_{2_{p'}}$}
    \State {\bf MPI send} $\{|\widetilde{\psi}^{B^1B^2...B^n}\rangle\}_{p'}$ to master processor.
\EndIf
\State \textbf{return} $|\widetilde{\psi}^{B^1B^2...B^n}\rangle$
\EndProcedure
\Statex
\Procedure{Action\_Smolyak\_basis}{$\hat{O}$, $\{|\psi^{B^1_{l_1}B^2_{l_2}...B^n_{l_n}}\rangle\}_p$, $i_1^{st}$, $i_2^{st}$}
  \For {$i^{st}$=$i_1^{st}$, $i_2^{st}$} 
    \State $\{l_1l_2...l_n\}\gets i^{st}$
    \State $ |\psi^{B^1_{l_1}B^2_{l_2}...B^n_{l_n}}\rangle \gets  \{|\psi^{B^1_{l_1}B^2_{l_2}...B^n_{l_n}}\rangle\}_p$
    \State $|\psi^{G^1_{l_1}G^2_{l_2}...G^n_{l_n}}\rangle  \gets  |\psi^{B^1_{l_1}B^2_{l_2}...B^n_{l_n}}\rangle$ 
    \State $|\widetilde{\psi}^{G^1_{l_1}G^2_{l_2}...G^n_{l_n}}\rangle=\hat{O}|\psi^{G^1_{l_1}G^2_{l_2}...G^n_{l_n}}\rangle$  
   \State $|\widetilde{\psi}^{B^1_{l_1}B^2_{l_2}...B^n_{l_n}}\rangle\gets |\widetilde{\psi}^{G^1_{l_1}G^2_{l_2}...G^n_{l_n}}\rangle$ 
   \State $\{|\widetilde{\psi}^{B^1_{l_1}B^2_{l_2}...B^n_{l_n}}\rangle\}_p\gets|\widetilde{\psi}^{B^1_{l_1}B^2_{l_2}...B^n_{l_n}}\rangle$
    \EndFor
  \State \textbf{return} $\{|\widetilde{\psi}^{B^1_{l_1}B^2_{l_2}...B^n_{l_n}}\rangle\}_p$
\EndProcedure
\end{algorithmic}
\end{algorithm}

\begin{figure}[ht!]
\begin{center}
\centering
\includegraphics[width=0.860\textwidth]{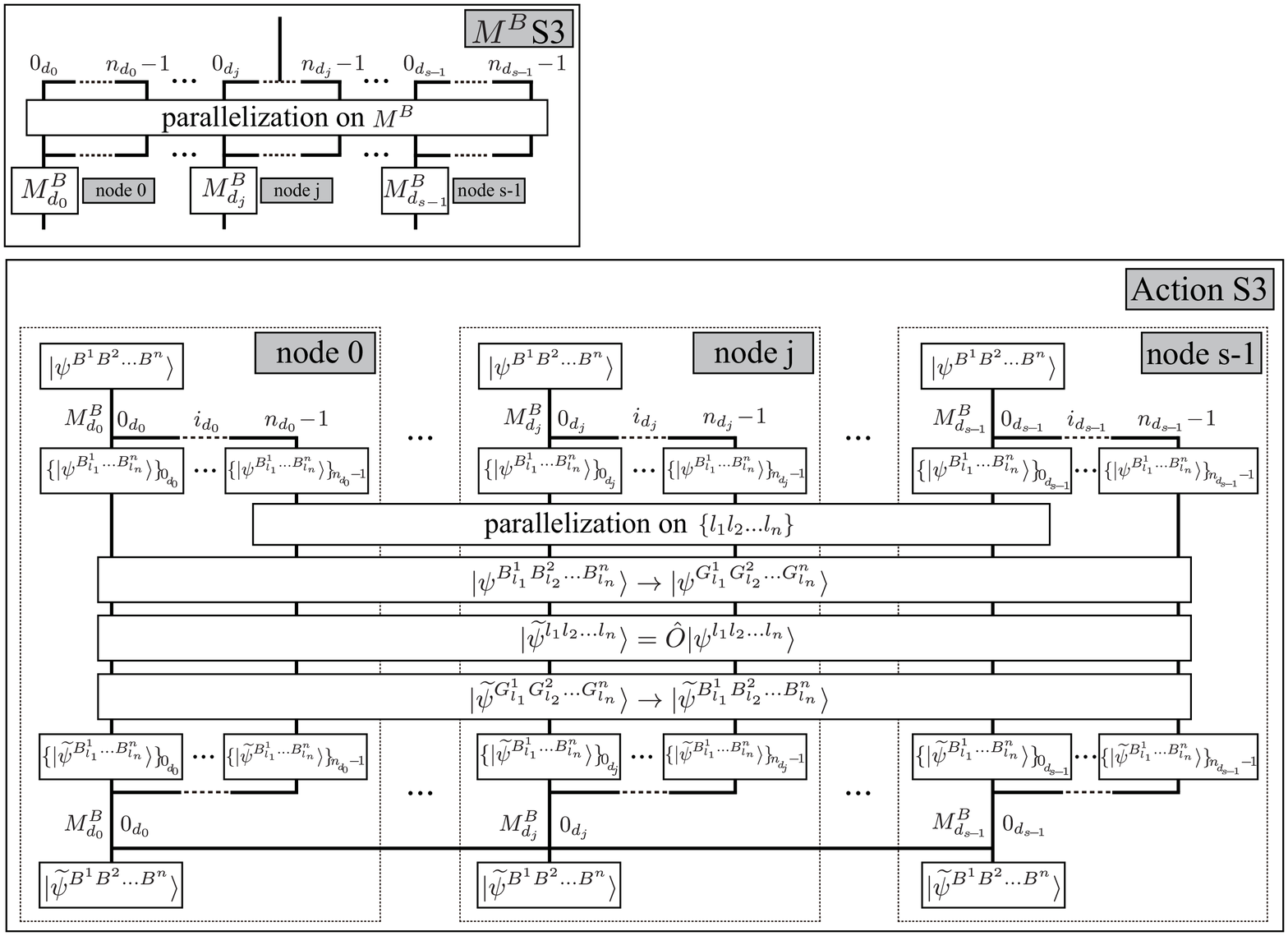}
\caption[]{Same as Fig.\ref{fig_action_S1} but for scheme 3.}\label{fig_action_S3}
\end{center}
\end{figure}

For large multi-node cluster, as a balance of the efficiency and memory usage, a combination of the two basic schemes leads to the third MPI scheme. In this scheme,  the available processors are divided into several groups. Scheme 2 is embedded in each group, while the works among groups are performed similarly to that of scheme 1. The $M^B$ are divided and assigned to a master processor in each group as $M^B_{d_i}$, where $d_i, i\in[0,s-1]$ is the group index, $s$ is the number of groups available.   The group is normally the cluster node. The relevant flowchart is shown in Fig\ref{fig_action_S3}. 

Practically, these schemes are further dressed. For instance, in Alg.\ref{algorithm_smolyak_MPI2},  the Smolyak terms should be packed to be sent to the other processor to reduce MPI communication. Meanwhile, due to the limited memory, in schemes 2 and 3, the Smolyak terms, in principle, should not be present at the same time to exhaust the memory. Therefore, an extra division should be implemented to perform the distribution of Smolyak terms several times, which could be setup in the simulation according to the available memory. 
Furthermore, according to the different sizes of Smolyak terms, the program efficiency could be further improved with the balance of the Smolyak terms assigned to each processor. We implement an auto-adjust mechanism to balance the works on each processor in scheme 2, based on the calculation and the communication time used on each processor in the previous step.



\begin{figure}[ht!]
\begin{center}
\centering
\includegraphics[width=0.750\textwidth]{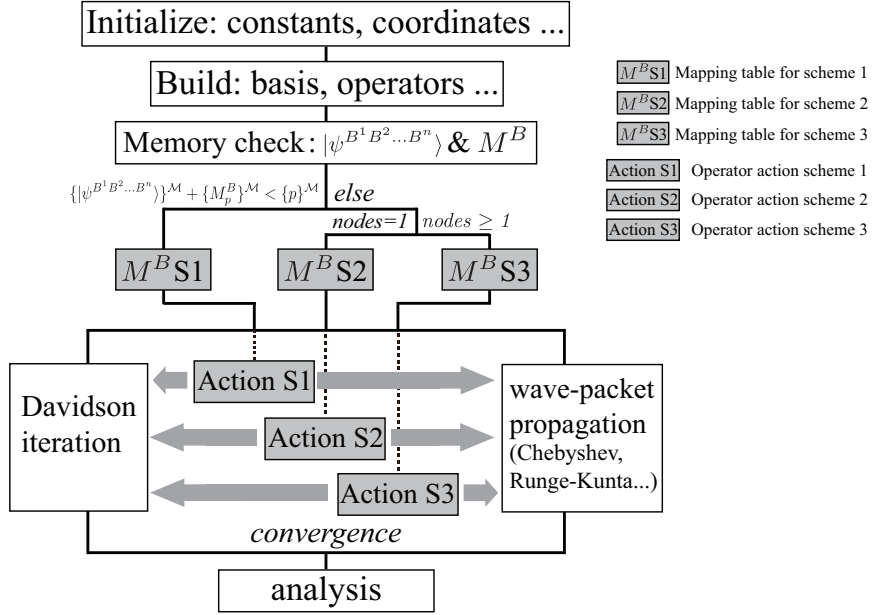}
\caption[]{The main flowchart of the code. See the main text for more details of the notations. }\label{fig_illustrator}
\end{center}
\end{figure}

These scheme are then implemented in the program \textsc{ElVibRot-MPI}~\cite{David2010_ElVibRot_Tnum}. A coarse flowchart of \textsc{ElVibRot-MPI} is shown in Fig.\ref{fig_illustrator}. The simulation is initialized with the physical constants and coordinates. Then the basis and operators are built according to the system considered. We apply the curvilinear coordinates for the simulation, of which the kinetic operator are obtained by \textsc{Tnum-Tana} package~\cite{David2002_curvilinear_coordinates}. A full manual of  \textsc{Tnum-Tana} 
is included in the code package. The follow on memory check will decide the MPI scheme chosen for the simulation if it is not specified in the input parameters. The strategy is as follows. Scheme 1 will be the first choice when there is enough memory assigned to each processor. Otherwise, we choose scheme 2 or 3 depends on the cluster nodes available for the simulation. Scheme 2 provides further options to save memory. The calculation and assignment of the mapping table are then decided according to the chosen scheme.  The program will when direct to the time-independent or -dependent simulation using the chosen operator action scheme. The main method employed for converging the quantum states of the system is the Davidson method, while the available methods for wave-packet propagation includes the Chebyshev method, the n-order Runge-Kunta method, the n-order Taylor expansion method, the short iterative Lanczos method, and the Bulirsch-Stoer method, etc.  The analysis is performed afterwards.

\setlist[description]{leftmargin=\parindent,labelindent=\parindent}

\section{Code usage}
A \textsc{ElVibRot} manual for a quick start of the program is prepared in the code package. It consists of a brief introduction of the program, the main input parameters, the installation, running of the code, and typical examples. 

\subsection{Parameters}
\textsc{ElVibRot} provides a highly customized simulation for different molecules with a series of parameters. The input parameters for the code takes the form of the ``namelist" in Fortran. It contains four main parts as follows.
\begin{description}[nolistsep]
\item [SYSTEM and CONSTANTS] define general parameters for parallelization, printing levels, energy unit, physical constants, etc. The available namelist includes:
\begin{lstlisting}[style=bash]
&system, &constantes
\end{lstlisting}
\item [COORDINATES] defines the curvilinear coordinates, the coordinates transformations and some aspects of the physical models (e.g. constraints). This section is a part of \textsc{Tnum}, see \textsc{Tnum-Tana} manual for details. The available namelist includes:
\begin{lstlisting}[style=bash]
&variables, &coord_transfo, &minimum
\end{lstlisting}
\item [OPERATORS and BASIS SETS] defines parameters of scalar operators (e.g. potential, dipole moments) and the contracted active and inactive basis sets. The available namelist includes:
\begin{lstlisting}[style=bash]
&basisnD, &inactive, &active
\end{lstlisting}
\item [ANALYSIS] defines parameters for time-dependent (including optimal control) or -independent calculations, intensities. The available namelist includes:
\begin{lstlisting}[style=bash]
&analyse, &davidson, &propa
\end{lstlisting}
\end{description}
For more details, see the \textsc{ElVibRot} quick manual.

\subsection{Installation}
The program could be run on general Linux and OS X platform. It could be compiled with {\it gfortran}, {\it  ifort}, and {\it  pgf90}. It supports to run with openMP when compiled with {\it gfortran} or {\it ifort}. To run with MPI, we should compile it with {\it mpifort}, which requires the installation of openMPI (V2.0 \& above). 

To compile the program, we need to modify the ``makefile".  There are several options. The main ones are as follows.
{\footnotesize
\begin{description}[nolistsep]
\item [{\bf F90}:] default \textbf{gfortran}; the compiler to use. Options includes ifort, pgf90 and mpifort.
\item [{\bf OPT}:] default \textbf{1}; the compiler optimization. \textbf{0} and \textbf{1} denotes turn-off or on the optimization, respectively.
\item [{\bf OMP}:] compilation with (\textbf{1}) or without (\textbf{0}) OpenMP. It will be automatically disabled when using compiler  {\it mpifort}.
\item [{\bf INT}:] default \textbf{4}; this enables to change the integer kind during the compilation to a ``long integer" (INT=8). This is useful for large calculations with Smolyak method. To run with MPI, the openMPI should be compiled with the same ``integer" accordingly.
\item [{\bf LAPACK}:] default \textbf{1} (with LAPACK); when  {\it LAPACK}=\textbf{1}, it enables the use of BLAS and LAPACK libraries. Otherwise ({\it LAPACK}=\textbf{0}), they are disabled.
\end{description}
}

Once compiled with ``make", we could use ``make UT" and ``make clean\_UT" to perform a unit test or clean the test results, respectively. We can also run MPI examples with ``make example" and clean the examples with ``make clean\_example".  The examples include the simulation of the pyrazine model~\cite{Worth1998_Pyrazine24D} with 12 or 24 degrees of freedom for the time-dependent propagation, the malonaldehyde with 21 degrees of freedom and the Henon-Heiles model with 6 degrees of freedom for the time-independent vibrational spectra calculation. Each example is set to test for different MPI schemes. 

\subsection{Run program}
To run the program, we should have the system information prepared (e.g. the coordinates, potential energy surface, etc. ).  The namelist could be directly a shell input, for instance,  

\begin{lstlisting}[style=bash]
$ vib << ** > output
 &system ... /
 &constantes ... /
 ... ... 
**
\end{lstlisting}
or prepared in a file named ``namlist" as 
\begin{lstlisting}[style=my_Common_list]
 &system ... /
 &constantes ... /
 ... ... 
\end{lstlisting}
and run with 
\begin{lstlisting}[style=bash]
$ vib > output
\end{lstlisting}
where ``vib" is the compiled executable file. To run with MPI, the second way is mandatory:
\begin{lstlisting}[style=bash]
$ mpirun -np number_of_processors vib > output
\end{lstlisting}
Moreover, we can use 
\begin{lstlisting}[style=bash]
$ mpirun -np number_of_processors MPI_output_control vib > output
\end{lstlisting}
 to further control the output of different processors with the file ``MPI\_output\_control", which is involved in the examples for MPI simulation.

\subsection{Example}
Here we provide an example for the calculation of the 28 lower eigen-states of 6D Henon-Heiles Hamiltonian with MPI:
\begin{eqnarray}
H=\frac{1}{2} \sum_{\kappa=1}^{f}\left(-\frac{\partial^{2}}{\partial q_{\kappa}^{2}}+q_{\kappa}^{2}\right)+\lambda \sum_{\kappa=1}^{f-1}\left(q_{\kappa}^{2} q_{\kappa+1}-\frac{1}{3} q_{\kappa+1}^{3}\right),~ (f=6)
\end{eqnarray}
It can be found in the examples provided for the program test. First, we prepare a bash file ``shell\_run" as:

\begin{lstlisting}[style=bash]
#!/bin/bash
## set up environment
num_cores=$1    ## number of parallelization processors
DIR_vib=$(directory_to_vib).
here=`pwd`

## "sub_system_HenonHeiles.f": file provides potentials and dipolar matrices used for Henon-Heiles Hamiltonian
cp sub_system_HenonHeiles.f $DIR_vib/sub_pot/sub_system.f  
cd $DIR_vib
make     ## compile program with "sub_system_HenonHeiles.f" 

## set up parameters
L=$2            ## parameters for Smolyak algorithm
LB=$L
LG=$L
B=2
nb_ana=28       ## number of target levels
dav=$3          ## if using Davidson iteration 
name_WP=$here/file_WPspectral ## wave function output name

## create namelist file, see manual for the details of parameters
cat > namelist << EOF
&system EVR=t MPI_scheme=3 /
&constantes ene_unit='au' /
&variables nrho=2 Without_Rot=t Centered_ON_CoM=f Gcte=t nb_Qtransfo=2 /
&Coord_transfo name_transfo='zmat' nat=5 /
 X
 X 1
 X 1 2
 1. 0 0 0  
 1. 0 0 0
&Coord_transfo name_transfo='active' /
 0 0 0  1 1 1  1 1 1
&minimum pot0=0.d0 nb_scalar_Op=0 read_Qsym0=t unit='bohr' /
 1.
 1.
 1.57
 0
 0
 0
 0
 0
 0
&basis_nD nb_basis=6 name="direct_prod" L_SparseBasis=$LB L_SparseGrid=$LG SparseGrid_type=4 /
 &basis_nD iQdyn=4 name="Hm" Q0=0. scaleQ=1. L_TO_nq_A=1 L_TO_nq_B=$B /
 &basis_nD iQdyn=5 name="Hm" Q0=0. scaleQ=1. L_TO_nq_A=1 L_TO_nq_B=$B /
 &basis_nD iQdyn=6 name="Hm" Q0=0. scaleQ=1. L_TO_nq_A=1 L_TO_nq_B=$B /
 &basis_nD iQdyn=7 name="Hm" Q0=0. scaleQ=1. L_TO_nq_A=1 L_TO_nq_B=$B /
 &basis_nD iQdyn=8 name="Hm" Q0=0. scaleQ=1. L_TO_nq_A=1 L_TO_nq_B=$B /
 &basis_nD iQdyn=9 name="Hm" Q0=0. scaleQ=1. L_TO_nq_A=1 L_TO_nq_B=$B /
&inactives /
&actives  test=f direct=4 direct_KEO=t Save_MemGrid=t direct_ScalOp=t /
&analyse  max_ana=$nb_ana max_ene=7. 'au' 
          name_file_spectralWP='$name_WP' davidson=$dav /
&davidson nb_WP=$nb_ana max_it=200 max_WP=5000 num_resetH=10
          lower_states=t conv_resi=0.0005 'au' conv_ene=0.0001 'au' 
          read_WP=t name_file_readWP='$name_WP' NewVec_type=4 /
EOF

## run program
mpirun -np $num_cores $here/MPI_output_control $DIR_vib/vib.exe > $here/output"_"$num_cores"MPIcores"

echo 'script finished'
\end{lstlisting}
Assuming we have 4 processors, by running 
\begin{lstlisting}[style=bash]
./shell_run 4 2 f
\end{lstlisting}
we could obtain a guess of the wave function by a direct diagonalization at low $L$. Then we can run formally with Davidson iteration by
\begin{lstlisting}[style=bash]
./shell_run 4 5 t
\end{lstlisting}
to obtain the target states. The obtained energy levels could be find under keywords ``Energy level" in the output file.

\section{Results}
\begin{figure}   [ht!]
\begin{center}
\centering
\includegraphics[width=0.800\textwidth]{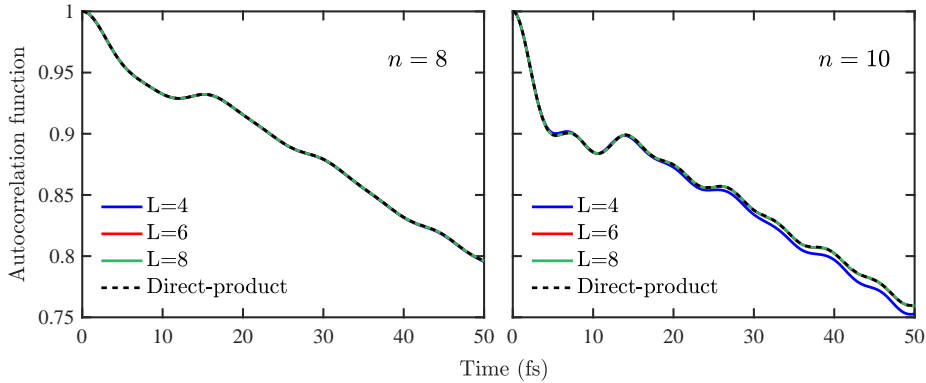}
\caption[]{The autocorrelation of the propagation of a Gaussian-packet in a quadratic potential obtained at different $L$ using \textsc{ElVibRot}-MPI. The smulation is perfromed for system of 8 or 10 degrees of freedom as indicated in the panels. The result is compared with that of direct-product scheme.  }\label{fig_autocorrelation}
\end{center}
\end{figure}

We first present a benchmark comparison of direct-product scheme and \textsc{ElVibRot}-MPI for the propagation of Gaussian-packet in a quadratic potential. The initial Gaussian-packet is properly located to ensure it is covered by Smolayk basis. This case can also be found in the program package. In Fig.\ref{fig_autocorrelation} we show the autocorrelation of the propagation up to 50 fs for the system of 8 or 10 degrees of freedom as indicated. There is already very good agreement at $L$=4 and $L=6$ for 8 and 10  degrees of freedom system, respectively. The simulation time for 10 degrees of freedom at $L$=6 is around 10 min with 12 processors using \textsc{ElVibRot}-MPI, which for direct-product scheme takes more than 65 hours.

\begin{figure}   [ht!]
\begin{center}
\centering
\includegraphics[width=0.80\textwidth]{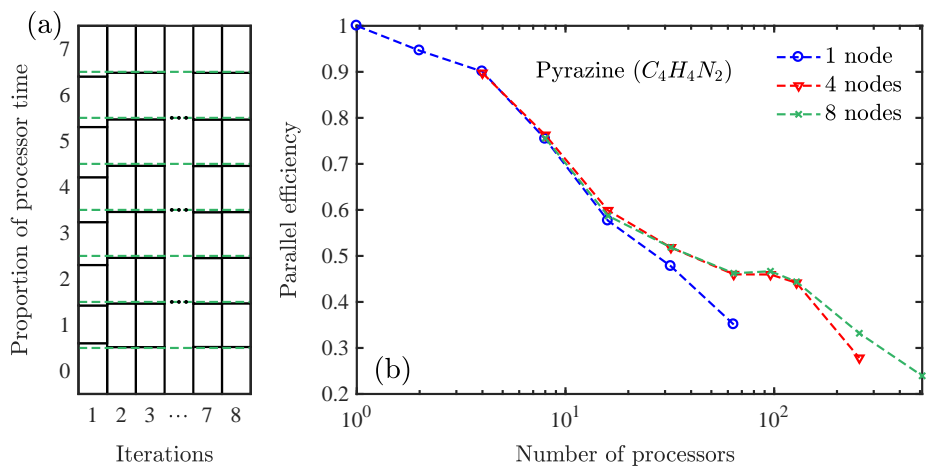}
\caption[]{A test with the simulation of Pyrazine (12 degrees of freedom) using \textsc{ElVibRot}-MPI. Panel (a) shows the auto-balance of Smolyak terms on 8 different processors in scheme 2. The green dashed lines mark each 1/8 proportion; Panel (b) presents the parallel efficiency of the simulation on 1, 4, or 8 nodes.
}\label{fig_result}
\end{center}
\end{figure}

Furthermore, we perform a efficiency test for \textsc{ElVibRot}-MPI using the propagation of Pyrazine vibronic model (12 degrees of freedom) ~\cite{Worth1998_Pyrazine24D}:
\begin{eqnarray}
H=\left(
\begin{array}{cc}
-\Delta & 0 \\
0 & \Delta
\end{array}\right)
+I\sum^n_{i=1}\frac{\omega_i}{2}({\bf p}_i^2+Q_i^2)
+\left(
\begin{array}{cc}
0 & \lambda Q_1 \\
\lambda Q_1 & 0
\end{array}\right)
+\sum^n_{i=2}
\left(
\begin{array}{cc}
k_i^1 & 0 \\
0 & k_i^2
\end{array}\right)Q_i.
\end{eqnarray}
The setup of the simulation could be found in the examples in the code package. The initial wave-packet is chosen as the first vibrational state of the model and propagates for 100 fs with the Chebyshev method in a time step $\Delta t=$0.1 fs.  
The simulation is performed with 1, 4, or 8 nodes to test the improvement as shown in Fig.\ref{fig_result}.
In panel (a) we present the auto-balance of Smolyak terms distributed to 8 processors during 8 iterations in scheme 2. The works on different processors are well-balanced after the first iteration. In Panel (b) we show the efficiency of the simulation as a function of available processors when different numbers of nodes (1, 4, or 8) get involved. 
Due to the relatively large MPI communications, the improvement is good. Moreover, when more nodes are available, the speedup would be even better, though the relatively poor communication between nodes. It comes from the reduction of mapping table size on each node as shown in the right panels of Fig.\ref{fig_illustrator} and the increasing of available memory. 

\section{Conclusion}
In a summary, we introduced a parallelized quantum dynamics simulation package \textsc{ElVibRot}-MPI. The program feathers the MPI implementation of Smolyak method to adapt the requirement of different machine. It shows a good MPI parallelization efficiency. \textsc{ElVibRot}-MPI makes possible the simulation of general molecules up to a few tens degrees of freedom without the limitation of the Hamiltonian.  A wide variety of parameters is provided, allowing a high flexibility setup of the simulation to adapt to different problems. Typically, it is applied for the calculation of vibrational levels, intensities for floppy molecular systems, the wave-packet propagation, and the quantum gate, etc.

\section*{Acknowledgments}
A.C. gratefully acknowledges the funding support from E-CAM European Centre of Excellence,  European Union's Horizon 2020 research and innovation program under Grant No. 676531.
We acknowledges the computation resource of the {\it styx} in
Institut de Chimie Physique, Universit\'{e} Paris-Saclay, and the {\it JUWELS} in J\"{u}lich Supercomputing Centre provided by Dr. Alan O'Cais in E-CAM.





\bibliography{cite}
\bibliographystyle{abbrv}







\end{document}